\documentclass[prb,aps, babel,twocolumn]{revtex4}
\usepackage{amsmath,amssymb,amstext,amscd,amsthm,amsopn,graphicx,indentfirst,mathrsfs,natbib}
\newcommand{\Jpar}{J_{\parallel}}
\newcommand{\Jper}{J_{\perp}}

\newcommand{\f}{\phi}

\renewcommand{\vec}[1]{{\bf #1}}

\graphicspath{{.}{./EPS/}}
\begin{document}

\title{Plaquette Order in a Dimerized Frustrated Spin-Ladder}

\author{Ofer Shlagman}
%\email{ofer.shlagman@live.biu.ac.il}

\affiliation{Department of Physics, Bar Ilan University, Ramat-Gan 52900, Israel}
\author{Efrat Shimshoni}
%\email{shimshe@mail.biu.ac.il}

\affiliation{Department of Physics, Bar Ilan University, Ramat-Gan 52900, Israel}
\begin{abstract}
We study the effect of dimerization (due to, e.g., spin-Peierls instability) on the
phase-diagram of a frustrated antiferromagnetic spin-1/2 ladder,
with weak transverse and diagonal rung coupling. Our analysis
focuses on a one-dimensional version of the model (i.e. a single
two-leg ladder) where we consider two forms of dimerization on the legs: columnar dimers (CD) and staggered
dimers (SD). We particularly examine the regime of parameters
(corresponding to an intermediate XXZ anisotropy) where the leg-dimerization and the
rung coupling terms are equally relevant. In both the CD and SD
cases we find that the effective field theory describing the
system is a self-dual sine-Gordon model, which favors ordering and
the opening of a gap to excitations. The order parameter, which
reflects the interplay between the leg and rung dimerization interactions,
represents a crystal of 4-spin plaquettes on which longitudinal
and transverse dimers are in a coherent superposition. Depending on the leg
dimerization mode these plaquettes are closed or open, however both
types spontaneously break reflection symmetry across the ladder.
The closed plaquettes are stable, while the open plaquette-order
is relatively fragile and the corresponding gap may be tuned to
zero under extreme conditions. We further find that a first order
transition occurs from the Plaquette order to a valence bond
crystal (VBC) of dimers on the legs. It is suggestive that in a
higher dimensional version of this system, this variety of
distinct VBC states with comparable energies leads to the
formation of domains. Effectively one-dimensional gapless spinon
modes on domain boundaries can possibly account for the
experimental observation of a spin-liquid behavior in a physical
realization of the model.
\end{abstract}

\pacs{75.10.Pq,75.10.Jm,75.30.Kz}

\maketitle
\section{Introduction}
Low dimensional quantum magnets attract a lot of experimental and
theoretical attention, due to the rich physics arising from their
enhanced quantum fluctuations, and competing interactions which
often induce nonclassical ground states. Most prominently, quantum
effects are manifested by spin-$\frac{1}{2}$ systems at one
dimension (1D).  The simplest model for 1D quantum
antiferromagnets is the XXZ Hamiltonian, describing a
spin-$\frac{1}{2}$ chain with nearest neighbor
interactions\cite{bethe},
 \begin{equation}
H=\sum_iJ_{xy}(S_{i+1}^{x}S_{i}^{x}+S_{i+1}^{y}S_{i}^{y})+J_z\sum_iS_{i+1}^{z}S_{i}^{z}
\label{eq:xxz}
\end{equation}
where $J_{\alpha}>0$ corresponds to antiferromagnetic exchange
interaction. The isotropic case $J_{xy}=J_z$ yields the 1D
Heisenberg model. This system has a nonclassical ground state at
$T=0$ which is a gapless liquid, characterized by a lack of
long range order and power law decay of the spin-spin
correlations - namely, a critical state. The properties of this 1D liquid state at low
temperatures can be evaluated in terms of gapless
spin-$\frac{1}{2}$ excitations named spinons, which can be
represented as interacting spinless Fermions and form a Luttinger
liquid.

 \begin{figure}[h]
\begin{center}
\includegraphics[height=125pt,width=250pt]{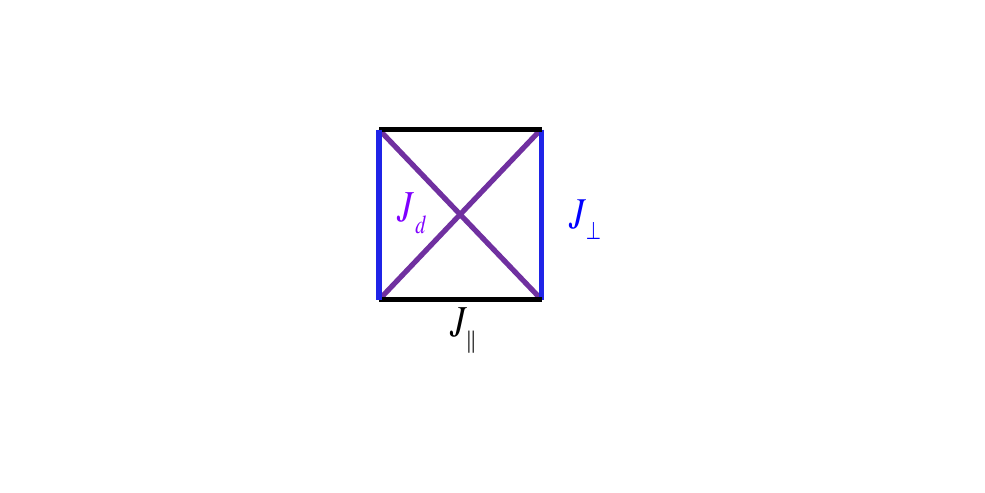}
\caption{(color online) schematic representation of the exchange
interactions of the NT model.} \label{fig:conf exch}
\end{center}
\end{figure}

Whether an analogous spin-liquid state can also be found in higher
dimensions, and under what conditions, is an important question
\cite{balents-2010}. Typically, the 1D Luttinger liquid state is
unstable to interchain couplings which tend to favor various types
of long-range order \cite{Dagotto1,Rice1,Rice2,Schulz,Essler}.
Therefore, a necessary condition is the presence of frustration
resulting from conflict between competing interactions. A
particular interesting model for frustrated spin-systems, on a 2D
cubic lattice, was introduced by Nersesyan and Tsvelik
\cite{ners-2003} (NT) as a possible realization of the long-sought
resonating valence bond (RVB) state\cite{anderson}. The model is
described by the Hamiltonian
\begin{widetext}
\begin{equation}
\label{eq: conf ham}
H=\sum_{j,\nu}\{J_{\parallel}\vec{S}_{j,\nu}\cdot\vec{S}_{j+1,\nu}+\sum_{\mu=\pm 1}[J_{\perp}\vec{S}_{j,\nu}+J_d(\vec{S}_{j+1,\nu}+\vec{S}_{j-1,\nu})]\cdot\vec{S}_{j,\nu+\mu}\},
\end{equation}
\end{widetext}
where $\nu$ enumerates chains and $j$ is the site number, $\Jpar$
is the intrachain exchange constant which couples neighboring
spins on the same chain, and $\Jper,J_d$ are the transverse and
diagonal interchain exchange constants respectively. The
interactions of this model are presented in Fig. \ref{fig:conf
exch}. The competition between exchange interactions on each
triangle prevents antiferromagnetic ordering. The model is
particularly interesting for the maximally frustrated special
ratio $J_{\perp}/J_d=2$, which corresponds to a critical point
between two phases of valence bond crystal (VBC)\cite{tsvelik2004}; i.e. a state
where pairs of spins form singlets (valence bonds) which are
localized, thus forming an ordered crystal. In the anisotropic
limit ($J_{\parallel}\gg J_{\perp}=2J_d$) of weakly coupled
chains, the ground state at this critical point was first argued
to be an RVB state\cite{ners-2003,moukouri-2004}: a state where
valence bonds undergo quantum fluctuations. The ground state is
then a superposition of different partitionings of spins into
valence bonds with no preference for any specific valence bond.
However, since then it was argued that the ground state could
still be a VBC \cite{starykh-2004,starykh-2010,Sindz}.

Recently, an experimental group has measured the thermodynamic properties of the material $\rm(NO)[Cu(NO_3)_3]$ (NOCuNO) \cite{vasiliev-2010}, which appears to be a good realization of the NT model in the weak coupling regime ($J_{\perp}\ll J_{\parallel}$). NOCuNO has the unique feature that due to the symmetry of the crystal structure its exchange interactions obey the special ratio $J_{\perp}=2J_d$. Hence, it provides a suggestive realization of the model exactly in the quantum critical point predicted in  Refs. [\onlinecite{ners-2003,starykh-2004}]. The contribution to the specific heat from magnetic excitations was fitted with an empirical formula which includes a term linear in $T$, characteristic to gapless spinons. Susceptibility and ESR measurements gave no indication for long range order throughout the whole measured temperature range, but indicate a considerable reduction compared to a standard spin-chain system at low $T$. The experimental data appear to indicate the existance of a spin-liquid component, that constitutes a fraction of the degrees of freedom in the system.

A more recent study of NOCuNO by Raman scattering
\cite{vasiliev-2012} indicated that a dynamical interplay between
spin and lattice degrees of freedom exists in this material which
might lead to novel phases. Moreover,  the Debye temperature in
NOCuNO is of the same order of magnitude as the spin exchange
interactions. It is known that when these two energy scales are
comparable, the spin-lattice coupling is
enhanced\cite{shimsh-2003,shlag-2012}. These observations motivate
the study of the effect of spin-lattice coupling, which was not
considered in earlier theoretical studies of the NT model.

One of the prominent consequences of spin-phonon coupling is the
emergence of spin-Peierls (SP) instability. This effect occurs
when the exchange couplings are modulated due to distortions in
the distance between neighboring atoms, yielding an alternation of
strong and weak bonds. The SP instability tends to dimerize the
spin-chain, therefore it can destroy the gapless liquid state and form a
VBC of longitudinal dimers\cite{affleck86,affleck87}. Away from the critical point of Ref.
[\onlinecite{ners-2003}], ($\Jper=2J_d$), a competition arises
between the SP instability (which tends to create longitudinal
dimers) and the transverse exchange coupling (which tends to
create transverse dimers). This competition may lead to a phase
transition between different types of dimer crystals, or induce a
new phase.

In one of the earlier theory works on the NT model \cite{starykh-2004}, Starykh and Balents
considered a dimer-dimer interaction term which may be generated
by higher-order interchain interactions, and studied the influence
of this interaction term via a renormalization-group (RG)
approach. They showed that various ordered nonmagnetic phases
(i.e. with zero magnetization) can form even in the absence of an explicit
dimerization term. The possible phases are two types of VBC, of staggered
and columnar longitudinal singlets, in addition to the rung
singlets and rung triplets (Haldane phase \cite{Haldane}) VBC
states predicted by NT. Later numerical studies
[\onlinecite{starykh-2010}] confirmed the emergence of such phases
for sufficiently strong interchain interactions. This suggests an
additional mechanism for dimerization besides interaction with
phonons.

\begin{figure}[b]
\begin{center}
\includegraphics[height=175pt,width=350pt]{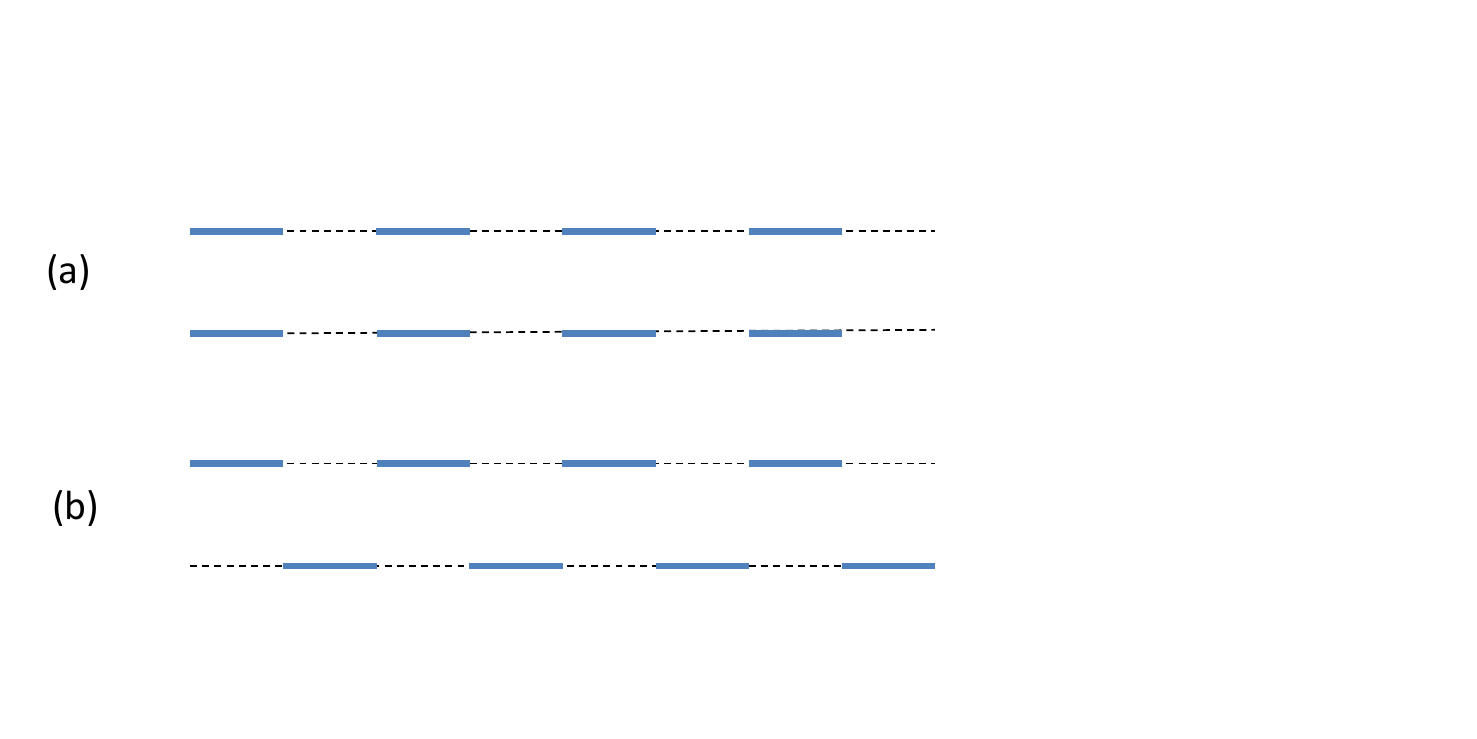}
\caption{(color online) Two possible ordering of valence bonds on
two noninteracting chains: (a) columnar dimers (CD), (b) staggered
dimers (SD). Thick red lines represent dimers on the stronger
bonds.} \label{fig:SDCD}
\end{center}
\end{figure}

A number of theoretical studies have also considered the effect of an explicit dimerization term on the Heisenberg spin-ladder without frustrating (diagonal) interactions\cite{affleck86,affleck87,sierra96,sierra98,cabra99,wang2000,carr2007}. Interestingly, these works found that although a dimerization term opens a gap when added to a  gapless spin-chain, adding such a term to a gapped spin ladder can lead to a gapless phase for suitably tuned dimerization and exchange couplings. It should be noted that
the possible phases crucially depend on the relative configuration of dimers on different legs of the ladder. Specifically in a two-leg ladder, there are two distinct configurations which
differ by the relative sign of the dimerization on the two
chains, and are dubbed columnar dimers (CD) and staggered
dimers (SD) [see Fig. \ref{fig:SDCD}]. The above mentioned works primarily examined the SD case in spin-1/2 Heisenberg ladders, which have been found \cite{sierra98,cabra99} to support massless spin excitations for sufficiently strong rung coupling. More recent studies considered the CD configuration along with the previously studied SD dimerization \cite{almeida2008,chitov2008,gibson2011}, finding that the ground state of the CD state is lower in energy and is always gapped.

In this paper, motivated by the experimental study of Ref. [\onlinecite{vasiliev-2012}], we examine the effect of leg-dimerization perturbations on the
low-energy physics of an anisotropic generalization of the NT model. For this purpose we consider a
two-leg ladder version of the NT model where we introduce dimerization terms on the legs as well as XXZ anisotropy of
all exchange couplings $\Jpar,\Jper,J_d$.
In distinction from the earlier studies \cite{affleck86,affleck87,sierra96,sierra98,cabra99,wang2000,carr2007,almeida2008,chitov2008,gibson2011} which consider the $SU(2)$-symmetric Heisenberg limit, we particularly focus on the case of an intermediate XXZ anisotropy ($\Jpar^z/\Jpar\sim 0.6$), where the rung and leg dimerization terms have an approximately equal scaling dimension, and hence strongly compete.
We consider both dimerization patterns (CD and SD) depicted in Fig. \ref{fig:SDCD}: assuming the dimerization to originate from a SP instability, the choice between them is dictated by the lattice deformation associated with the coupling to a certain phonon mode. Considering all the
interactions (perpendicular, diagonal and leg dimerization) in a Bosonization
description, we map the model onto an effective self dual
sine-Gordon model which we show is equivalent to a spin-chain in a
staggered and tilted magnetic field. Our main result is that the
effect of the leg dimerization on the NT model may lead to a
``Plaquette-ordered" state: a crystal of 4-spins plaquettes where
on each plaquette there is a coherent superposition of
longitudinal and transverse dimers, in a configuration which breaks reflection symmetry across the ladder. There are two types of
plaquette ground states corresponding to the two types of
dimerized patterns CD and SD. Generically, in both cases the ground state is
gapped. The tuning of parameters [the dimerization
$\delta\Jpar$, the inter-leg coupling ($\Jper-2J_d$) and the
anisotropies $J_\alpha^z/J_\alpha^{xy}$] leads to a smooth
interpolation between longitudinal and transverse VBC (each being recovered in the appropriate limit case), without a
second order phase transition.
Similarly to Chitov et. al.  (Ref. [\onlinecite{chitov2008}]), we find that the gap is larger in the CD case,
hence the plaquette order in this case is more stable. Moreover, for the SD case (Fig. \ref{fig:SDCD}(b)),
a 1D gapless (critical) state can
apparently be recovered under extreme conditions:
strong anisotropy of $\Jper$, $J_d$, and rung exchange of the
order of magnitude of the leg exchange. However, the closing of a
gap in the Plaquette-ordered state is always preempted by a {\it
first order} transition to the original VBC of leg-dimers.

The paper is organized as follows: in Sec. \ref{sec:model} we
derive the low-energy model for the spin system in terms of Bosonic fields. In Sec.
\ref{sec:phase_diagram} we analyze the model and demonstrate the
emergence of the Plaquette-order (subsection
\ref{sec:phase_diagram}A) and the phase transition to the VBC
state (subsection \ref{sec:phase_diagram}B). Technical details of the Fermionization method and of the
calculations of dimer correlation functions are discussed in
Appendices A and B, respectively. Finally, in Sec. \ref{sec:summary}, we summarize the
results and discuss their possible relevance to the behavior of
the 2D realizations of the frustrated coupled chains model.
\section{The Model}
\label{sec:model}
We consider a two-leg  XXZ ladder version of  the NT model (Eq. (\ref{eq: conf ham}))
\begin{widetext}
\begin{align}
\label{eq:single}
&H_{ladder}=\sum_j\left\{\sum_{\nu=1,2}\left\{\frac{\Jpar}{2}\left[S_{j,\nu}^+S_{j+1,\nu}^-+H.c\right]+\Jpar^zS_{j,\nu}^zS_{j+1,\nu}^z\right\}\right.\nonumber\\
&\left.+\frac{\Jper}{2}\left[S_{j,1}^+S_{j,2}^-+H.c\right]+\Jper^zS_{j,1}^zS_{j,2}^z+\frac{J_d}{2}[S_{j,1}^+(S_{j+1,2}^-+S_{j-1,2}^-)+H.c]+J_d^zS_{j,1}^z(S_{j+1,2}^z+S_{j-1,2}^z)\right\},
\end{align}
\end{widetext}
 with strong intrachain coupling ($\Jpar\gg\Jper,J_d$) and where all exchange couplings are positive. We then introduce a dimerization term along the legs of the ladder (e.g. due to spin-Peierls instability), described by a contribution to the Hamiltonian of the form
\begin{align}
\label{eq:Hp}
&H_P^{\sigma}=\frac{\delta \Jpar^{xy}}{2}\sum_j(-)^j[S_{j,1}^+S_{j+1,1}^-+\sigma S_{j,2}^+S_{j+1,2}^-+H.c]+\nonumber\\
&\delta\Jpar^z\sum_j(-)^j[S_{j,1}^zS_{j+1,1}^z+\sigma S_{j,2}^zS_{j+1,2}^z].
\end{align}
This term describes a static deformation of the exchange constants along the legs. In the absence of transverse coupling between the chains $1$ and $2$, the ground state of each chain is a product of singlets on the strong bonds (i.e., a one-dimensional VBC), with a gap of the order of $\delta\Jpar$ between the singlet ground state and the lowest triplet excitation. The parameter $\sigma=\pm$ is the relative sign of the dimerization on the two chains. The leg dimerization term is relevant as long as we are away from the ferromagnetic transition point, therefore, in our case, it is strongly relevant and tends to open a gap on each chain independently. As a result, in the case of two uncoupled chains there are two possible patterns in which a VBC can form on the chains, depending on $\sigma$: $\sigma=+$ yields a columnar dimers state (CD) and $\sigma=-$ a staggered dimers state (SD), as shown in Fig. \ref{fig:SDCD}.

To derive the low-energy model of the system, we first use the Jordan-Wigner transformation which maps the spin operators into Fermion fields. In the absence of magnetic field the Fermi energy is at the middle of the band, and the Fermion operators can be expressed in terms of Bosonic ones related to the Fermion density fluctuations:
\begin{align}
\label{eq:psi}
\psi_{R/L,\nu}=\frac{1}{\sqrt{2\pi a}}e^{-i(\pm\f_{\nu}-\theta_{\nu})},
\end{align}
where $R,L$ stands for right and left moving Fermions, respectively, $a$ is the lattice constant and $\nu=1,2$ is the leg index. This procedure can be shortly summarized by the following spin to Boson transformation
\begin{align}\label{eq:spin-boson}
&S^+_{\nu}(x)=\frac{e^{-i\theta_{\nu}(x)}}{\sqrt{2\pi a}}[(-)^x+\cos(2\phi_{\nu}(x))]\; ,\nonumber\\
&S^z_{\nu}(x)=-\frac{1}{\pi}\partial_x\phi_{\nu}(x)+\frac{(-)^x}{\pi a}\cos(2\phi_{\nu}(x))\;.
\end{align}
 We now Bosonize the Hamiltonian $H=H_{ladder}+H_P^{\sigma}$, starting from
  \begin{align}
  \label{Hladder}
  &H_{ladder}=\int dx\sum_{\nu=1,2}\frac{u}{2\pi}\left[\frac{1}{K}(\partial_x\f_{\nu})^2+K(\partial_x\theta_{\nu})^2\right]\nonumber\\
  &+\int dx\left[\frac{g}{(2\pi a)^2}\cos(\theta_1-\theta_2)+\frac{g^z}{(2\pi a)^2}\cos[2(\f_1-\f_2)]\right.\nonumber\\
  &\left.+\frac{g^z}{(2\pi a)^2}\cos[2(\f_1+\f_2)]\right]+(\Jper^z+2J_d^z)a\int dx\frac{\partial_x\f_1\partial_x\f_2}{\pi^2},\nonumber\\
  &g\equiv\frac{2\pi a(J_{\perp}-2J_d)}{(2\pi a)^2},\;g^z\equiv\frac{2\pi a(J_{\perp}^z-2J_d^z)}{(2\pi a)^2}.
  \end{align}
Here $u$ and $K$ are the Luttinger parameters of each chain on its own and are given by\cite{gia}
\begin{align}
\label{uK_def}
  &u=\frac{\Jpar}{2}\cdot\frac{\sqrt{1-(\Jpar^z/\Jpar)^2}}{1+\frac{1}{\pi}\arccos(-\Jpar^z/\Jpar)},\nonumber\\
  &K=\frac{\pi}{2\arccos(-\Jpar^z/\Jpar)}.
 \end{align}
Note that the frustrating interactions $J_d$ and $J_d^z$  enable the tuning of $g$ and $g^z$ independently of each other, and make them relatively small. This consequence of the frustration is important for our discussion since different phases may appear as a function of the ratio $g/g^z$.

Eq. (\ref{Hladder}) can be written more conveniently in terms of independent symmetric and antisymmetric modes\cite{gia,SNT,GNT}
 \begin{align}
 \f_{s/a}=\frac{\f_1\pm\f_2}{\sqrt{2}},\quad\theta_{s/a}=\frac{\theta_1\pm\theta_2}{\sqrt{2}}.
\label{eq:fas}
\end{align}
 Then, $H_{ladder}$ assumes the form
\begin{align}
\label{eq:Has}
&H_{ladder}=\int dx\left\{\sum_{\mu=a,s}\frac{u_{\mu}}{2\pi}\left[\frac{1}{K_{\mu}}(\partial_x\phi_{\mu})^2+K_{\mu}(\partial_x\theta_{\mu})^2\right]\right.\nonumber\\
&+g\cos(\sqrt{2}\theta_a)\nonumber\\
&\left.+g^z\cos(\sqrt{8}\phi_a)+g^z\cos(\sqrt{8}\phi_s)\right\}
\end{align}
 where for $\Jper,J_d\ll\Jpar$
 \begin{align}
 \label{eq:Kas}
  &K_{a,s}\cong K\left[1\pm\gamma\right],\nonumber\\
 &u_{a,s}\cong u\left[1\mp\gamma\right],\nonumber\\
 &\gamma\equiv\frac{K(J_{\perp}^z+2J_d^z)a}{2\pi u}.
 \end{align}
In ladders with antiferromagnetic legs where the XXZ anisotropy is anywhere in the range between the Heisenberg and XX limits (such that $1/2\leq K\leq 1$), both sectors $s$ and $a$ are gapped. The nature of the resulting phase depends on the sign of $g$, $g^z$: negative sign (i.e. effectively ferromagnetic rung coupling) yields the Haldane phase, while positive sign (effectively antiferromagnetic rung coupling) yields a crystal of rung singlets \cite{gia,SNT,GNT}.

Bosonization of $H_P^{\sigma}$ (see, e.g., Ref. [\onlinecite{gia}] for a detailed derivation) yields a term more conveniently written in terms of the original fields $\phi_1$, $\phi_2$:
\begin{align}
\label{eq:HP}
&H_P^{\sigma}\sim g_P\int dx[\sin(2\f_1)+\sigma\sin(2\f_2)],\nonumber\\
&g_P\equiv\frac{\delta\Jpar}{\pi a}.
\end{align}
Then, substituting Eq. (\ref{eq:fas}) into Eq. (\ref{eq:HP}), recasts $H_P^{\sigma}$ as a coupling term between the $a$ and $s$ sectors. The resulting full Hamiltonian is
 \begin{widetext}
\begin{align}
\label{eq:HNT}
&H=H_{ladder}+H_P^{\sigma}=H_a+H_s+H_{as}^{\sigma},\nonumber\\
&H_a=\frac{u_a}{2\pi}\int dx\left[\frac{1}{K_a}(\partial_x\f_a)^2+K_a(\partial_x\theta_a)^2\right]+g\int dx\cos(\sqrt{2}\theta_a)+g^z\int dx\cos(\sqrt{8}\f_a),\nonumber\\
&H_s=\frac{u_s}{2\pi}\int dx\left[\frac{1}{K_s}(\partial_x\f_s)^2+K_s(\partial_x\theta_s)^2\right]+g^z\int dx\cos(\sqrt{8}\f_s),\nonumber\\
&H_{as}^{\sigma}=g_P\int dx[\sin[\sqrt{2}(\f_s+\f_a)]+\sigma\sin[\sqrt{2}(\f_s-\f_a)]].
\end{align}
 \end{widetext}

 Note that at the critical point of the NT model, which for anisotropic rung coupling requires both $\Jper=2J_d$ and $\Jper^z=2J_d^z$, the coefficients $g,\;g^z$ in Eq. (\ref{eq:HNT}) vanish. However, away from the critical point we must consider the competition between all the cosine terms appearing. The relevance of the terms is determined by the scaling dimensions. Let us denote by $d_{a/s}^z$ the scaling dimensions of the terms $g^z\cos(\sqrt{8}\f_{a/s})$ respectively in the $a/s$ sector, $d$ of the term with coefficient $g$, and $d_P$ the scaling dimension of the term with coefficient $g_P$. Employing a perturbative RG, these scaling dimensions are given by
 \begin{align}
 \label{eq:scaling}
 d=\frac{1}{2K_a},\; d_a^z=2K_a,\;d_P=\frac{1}{2}(K_a+K_s),\;d_s^z=2K_s.
 \end{align}
Since $\gamma$ (Eq. (\ref{eq:Kas})) is positive, $d_s^z<d_a^z$. Therefore $g^z\cos(\sqrt{8}\f_a)$ is always the least relevant term, and will be neglected henceforth. Comparing $d_P$ and $d_s^z$ we find that for weak rung coupling ($\gamma<1/2$),
the $g_P$ term is also more relevant than the $g^z\cos(\sqrt{8}\f_s)$ term. This analysis of the scaling dimensions leads to the conclusion that for weak rung coupling the dominant terms which govern the low energy description are  the $g$ and $g_P$ terms, which indicate a potential competition between the leg-dimerization and the rung coupling (which tends to form rung-dimers for $g>0$). There is a special value of $K$ for which the scaling dimensions of the most relevant terms are equal:  from Eqs. (\ref{eq:Kas}), (\ref{eq:scaling}) we find $d=d_P$ for
 \begin{align}
 K^*=\frac{1}{\sqrt{2(1+\gamma)}}\cong\frac{1}{\sqrt{2}}.
\end{align}
For this value of $K$ (which corresponds to an intermediate anisotropy $\Jpar^z/\Jpar\approx 0.6$ - see Eq. (\ref{uK_def})) the competition between the leg and rung  dimerization terms is maximal. In the remaining paper we therefore focus our attention primarily on the regime of parameters where $K\sim K^*$.

We now recall that for arbitrary $K$ in the regime of interest $1/2\leq K< 1$ (i.e. $0<\Jpar^z\leq\Jpar$, and in particular for $K\sim K^*$), $K_s<1$. Hence the term $g^z\cos(\sqrt{8}\f_s)$ is also relevant, and tends to lock the value of the symmetric field $\f_s$. This affects the interaction term $H_{as}^{\sigma}$ which has a different form for SD ($\sigma=-$) and CD ($\sigma=+$) configurations:
 \begin{align}
 \label{eq:Has}
 H_{as}^+=2g_P\int dx\sin(\sqrt{2}\f_s)\cos(\sqrt{2}\f_a),\nonumber\\
 H_{as}^-=2g_P\int dx\cos(\sqrt{2}\f_s)\sin(\sqrt{2}\f_a).\nonumber\\
  \end{align}
In a semiclassical approximation, the $\cos(\sqrt{8}\phi_s)$ term in Eq. (\ref{eq:HNT}) obtains a finite expectation value which minimizes $H_s$. This depends on the sign of $g^z$: for $g^z>0$,
  \begin{align}
  \langle\cos(\sqrt{8}\phi_s)\rangle\cong-1\Rrightarrow \sqrt{2}\langle\phi_s\rangle\cong\pi/2
   \end{align}
while for $g^z<0$ (effectively ferromagnetic rung coupling), $\langle\f_s\rangle=0$.
We can therefore replace $\f_s$ with its expectation value everywhere it appears in the interaction term $H_{as}^{\sigma}$. Noting that a change in sign of $g^z$ is essentially equivalent to trading the roles of $\sigma=+$ and $\sigma=-$, we hereon confine ourselves to $g^z>0$: our final conclusions on the behavior dictated by the two distinct dimerization patterns will be exchanged in the case $g^z<0$.

For $\sigma=+$, in this semi-classical approach we obtain an effective Hamiltonian for the anti-symmetric mode $\phi_a$
\begin{widetext}
\begin{align}
\label{eq:Ha_eff}
&H_a^{eff}=\int dx\left\{\frac{u_{a}}{2\pi}[\frac{1}{K_{a}}(\partial_x\phi_{a})^2+K_{a}(\partial_x\theta_{a})^2]+
g\cos(\sqrt{2}\theta_a)+2g_P\cos(\sqrt{2}\phi_a)\right\},\nonumber\\
\end{align}
\end{widetext}
where the last term results from the substitution $\sin(\sqrt{2}\f_s)\cong\sin(\sqrt{2}\langle\f_s\rangle)=1$  in $H_{as}^{+}$ [Eq.(\ref{eq:Has})].
$H_a^{eff}$ belongs to a class of self dual sine-Gordon models which have known solutions\cite{Lech}. This will be analyzed in detail in the next section, and will be shown to yield a gapped, plaquette-ordered ground state. However for $\sigma=-$, this naive semiclassical approximation would result in $H_{as}^-=0$, since $\cos(\sqrt{2}\f_s)\cong\cos(\sqrt{2}\langle\f_s\rangle)=0$. This would imply that the leg dimerization is completely suppressed, and by tuning $g$ to zero one recovers a gapless Luttinger liquid state in the antisymmetric sector. Since $H_P^{\sigma}=H_{as}^{\sigma}$ is strongly relevant, it seems improbable that it can vanish completely from the low energy theory; rather, this is an artefact of the naive assumption $\langle\cos(\sqrt{8}\f_s)\rangle=-1$. Quantum fluctuations generally induce a finite expectation value which may be different than $-1$, in which case $H_{as}^-$ does not vanish. In the next section we discuss its contribution more carefully.

\section{Phase Diagram}
\label{sec:phase_diagram}
In what follows we focus on the
properties of the model for $K\sim K^*\approx1/\sqrt{2}$, in
which, as noted in the previous section, the terms responsible for
the formation of leg and rung singlets are equally relevant. We
derive a general theory that accounts for both the SD and CD
configurations of the leg dimerization, given in terms of an
effective Hamiltonian similar in form to Eq. (\ref{eq:Ha_eff}). As
we show below, this effective model indicates the potential
formation of a ``Plaquette order" phase.
\bigskip

\subsection{Emergence of Plaquette Order}
As a first step in our analysis of the Hamiltonian Eq. (\ref{eq:HNT}), we
introduce an auxiliary $Z_2$ order-parameter field $\hat{\tau}$ which allows to explore the potential for spontaneous breaking of reflection symmetry across the ladder. When this occurs, this field acquires an expectation value $\tau=\pm 1$. As we show below, the value of $\tau$ dictates a broken symmetry ground state where dimerization on one leg of the ladder is stronger than the other. We then
define new Bosonic fields via the transformation
\begin{align}
\phi_{p,\tau}=\phi_s+\tau\phi_a,\;\theta_{\tau}=\tau\theta_a,\nonumber\\
\phi_{f,\tau}=\sqrt{2}\phi_s,\;\theta_{f,\tau}=\frac{1}{\sqrt{2}}(\theta_s-\tau\theta_a)
\label{eq:nonunitary}
\end{align}
which preserve the canonical commutation relations
\begin{align}
[\phi_{l,\tau}(x),\theta_{l',\tau}(x')]=i\pi \delta_{ll'}\text{sign}(x-x').
\end{align}
Note that from Eq. (\ref{eq:fas}), $\f_{p,\tau}$ and $\theta_{f,\tau}$ are simply related to the original fields $\f_{\nu},\;\theta_{\nu}$ on the isolated legs $\nu=1,2$: for $\tau=+$, $\f_{p,\tau}=\sqrt{2}\f_1$ and $\theta_{f,\tau}=\theta_2$, and for $\tau=-$ the roles of $1,2$ are interchanged. Substituting Eq. (\ref{eq:nonunitary}) in Eq. (\ref{eq:HNT}) and removing the least relevant term $\cos(\sqrt{8}\f_a)$, we get
\begin{widetext}
\begin{align}
\label{eq:Hpf}
&H=H_{p}^{\sigma}+H_f+H_{pf}^{\sigma}\nonumber\\
&H_{p}^{\sigma}=\frac{u_{p}}{2\pi}\int dx\left[\frac{1}{K_{p}}(\partial_x\phi_{p,\tau})^2+K_{p}(\partial_x\theta_{p,\tau})^2\right]+g\int dx\cos(\sqrt{2}\theta_{p,\tau})+g_P(\delta_{\tau,-\sigma}+\sigma\delta_{\tau,\sigma})\int dx\sin(\sqrt{2}\phi_{p,\tau}),\nonumber\\
&H_f=\frac{u_f}{2\pi}\int dx\left[\frac{1}{K_f}(\partial_x\phi_{f})^2+K_f(\partial_x\theta_{f})^2\right]+g^z\int dx\cos(2\phi_f),\\
&H_{pf}^{\sigma}=\int dx\left\{-\frac{\sqrt{2}u_a}{K_a}\partial_x\phi_{f}\partial_x\phi_{p,\tau}+2\sqrt{2}u_sK_s\partial_x\theta_{f}\partial_x\theta_{p,\tau}+
g_P(\delta_{\tau,\sigma}+\sigma\delta_{\tau,-\sigma})[\sin(2\phi_{f})\cos(\sqrt{2}\phi_{p,\tau})-
\cos(2\phi_{f})\sin(\sqrt{2}\phi_{p,\tau})]\right\}\nonumber
\end{align}
\end{widetext}
(for abbreviation, we removed the index $\tau$ on the $f$-sector fields, where it turns out to be of no significance).
Here the velocities are given by
\begin{align}
&u_{p}=u_a\sqrt{\frac{u_aK_a+u_sK_s}{u_aK_a}}\cong\sqrt{2}u(1-\gamma),\nonumber\\
&u_f=\sqrt{u_sK_s\left(\frac{u_a}{K_a}+\frac{u_s}{K_s}\right)}\cong\sqrt{2}u
\end{align}
and the Luttinger parameters are
\begin{align}
&K_{p}=\sqrt{\frac{K_a(u_aK_a+u_sK_s)}{u_a}}\cong\sqrt{2}K(1+\gamma),\nonumber\\
&K_f=\sqrt{\frac{4u_sK_s^2K_a}{u_aK_s+u_sK_a}}\cong\sqrt{2}K\; ,
\end{align}
where in the final approximations we neglect terms of order $\gamma^2$.
%In the vicinity of $K^*$ the terms $g\cos(\theta_{p,\tau}$ and $g^z\cos(2\f_{p,\tau})$ are relevant, and the density-density terms in $H_{inter}^{\sigma}$ at %most renormalize the coefficients.

Next we use the fact that for $K\cong1/\sqrt{2}$ such that $K_f\cong1$, the $f$ sector reduces to a model of gapped free Fermions. Employing Eq. (\ref{eq:psi}) (with $\f_{\nu},\theta_{\nu}$ replaced by $\f_f,\theta_f$), the cosine term in $H_f$ can be refermionized to give
\begin{align}
\label{eq:cosphi_ferm}
\cos(2\phi_f)=\pi a(\psi_R^{\dagger}\psi_L+\psi_L^{\dagger}\psi_R).
\end{align}
This term opens a gap to excitations given by
\begin{align}
\Delta=\pi ag^z
\end{align}
(see Appendix \ref{app:referm} for details).
For low $T\ll\Delta$ we can simplify the interaction term $H_{pf}^{\sigma}$ by a mean-field approximation. This amounts to replacing $\cos2\phi_f$ as well as $\partial_x\phi_f,\;\partial_x\theta_f$ and $\sin2\phi_f$ in $H_{pf}^\sigma$ [Eq. (\ref{eq:Hpf})] by their expectation values. To leading order in $\Delta a/u_f$, this yields
\begin{align}
\label{eq:expect}
\langle\sin2\phi_f\rangle=0,\;\langle\partial_x\phi_f\rangle=0,\langle\partial_x\theta_f\rangle=0\; ,
\end{align}
\begin{align}
\label{eq:Of}
O_f\equiv\langle\cos2\phi_f\rangle\sim-\frac{\vert\Delta\vert}{u_f/a}\ln\left[\frac{u_f/a}{\vert\Delta\vert}\right]\; .
\end{align}
Substituting back into Eq. (\ref{eq:Hpf}) we obtain an effective model for the $p$ sector:
\begin{align}
\label{eq:Hp}
&H_{p}^{\sigma,eff}= \frac{u_{p}}{2\pi}\int dx\left[\frac{1}{K_{p}}(\partial_x\phi_{p,\tau})^2+K_{p}(\partial_x\theta_{p,\tau})^2\right]\nonumber\\
&+g\int dx\cos(\sqrt{2}\theta_{p,\tau})+\tilde{g}_P(\sigma,\tau)\int dx\sin(\sqrt{2}\phi_{p,\tau}),\nonumber\\
&\tilde{g}_P(\sigma,\tau)\equiv(\delta_{\tau,-\sigma}+\sigma\delta_{\tau,\sigma}) g_P\left[1-\sigma O_f\right].
\end{align}

The leg dimers configuration on the ladder, CD ($\sigma=+$) or SD ($\sigma=-$) is encoded in the effective Hamiltonian
Eq. (\ref{eq:Hp}) by the parameter $\tilde{g}_P(\sigma,\tau)$. Its dependence on $\sigma$, $\tau$ reflects a crucial distinction between the two patterns: first, in the CD case this effective parameter is symmetric in $\tau$, $\tilde{g}_P(+,+)=\tilde{g}_P(+,-)$; i.e., the Hamiltonian is identical for $\tau=\pm$. In contrast, the SD configuration yields $\tilde{g}_P(-,+)=-\tilde{g}_P(-,-)$, namely two distinct effective Hamiltonians for $\tau=\pm$. Second, since $O_f<0$ [see Eq.(\ref{eq:Of})], $\tilde{g}_P(+,\tau)$ is always finite and obeys $|\tilde{g}_P(+,\tau)|>|\tilde{g}_P(-,\tau)|$. Most prominently, only in the SD case ($\sigma=-$) a gapless liquid phase can be reached. This occurs for very special values of the exchange interactions where {\it both} $g$ and $\tilde{g}_P(-,\tau)$ vanish: when $\Jper^{xy}=2J_d^{xy}$ (i.e., at the critical point of the NT model), and at the same time $(\Jper^z-2J_d^z)\sim\Jpar$ so that $O_f\sim1$ [\onlinecite{foot1}]. Then all the interaction terms vanish and we are left with a Luttinger-liquid model. For this extremely-fine-tuned point, this analysis gives a Luttinger liquid phase in the case where the dimerization on the legs of the ladder is of the SD type (Fig 2(b)). This Luttinger liquid then describes gapless spinons on a single chain composed of the interlaced chains 1 and 2.
It is essentially the same as the "snake-chain" described in Ref. [\onlinecite{cabra99}] for the isotropic Heisenberg  spin-ladder. We note, however, that under these conditions another ordered ground state is likely to be favored, as will be discussed in subsection B.

To explore the more generic case where $g$ and/or $\tilde{g}_P$ are finite, we next rescale the fields, $\tilde{\phi}_{p,\tau}=\frac{\phi_{p,\tau}}{\sqrt{2}},\;\tilde{\theta}_{p,\tau}=\sqrt{2}\theta_{p,\tau}$, and accordingly the Luttinger parameter $\tilde{K}_{p}=\frac{K_{p}}{2}$ to obtain
\begin{widetext}
\begin{align}
&H^\sigma_{eff}=\int dx \left\{\frac{u_{p}}{2\pi}\left[\frac{1}{\tilde{K}_{p}}(\partial_x\tilde{\phi}_{p,\tau})^2+\tilde{K}_{p}(\partial_x\tilde{\theta}_{p,\tau})^2\right]+
g\cos(\tilde{\theta}_{p,\tau})+\tilde{g}_P(\sigma,\tau)\sin(2\tilde{\phi}_{p,\tau})\right\}.
\end{align}
\end{widetext}
For $g>0$ the interaction terms $\cos(\tilde{\theta}_{p,\tau}),\;\sin(2\tilde{\phi}_{p,\tau})$ are dimerization operators, each one creates different dimers: the $\cos(\tilde{\theta}_{p,\tau})$ creates dimers along the rungs of the ladder, and $\sin(2\tilde{\phi}_{p,\tau})$ creates dimers along the legs of the ladder. It is convenient to define $\tilde{\phi}_{\tau}=\tilde{\phi}_{p,\tau}-\pi/4,$ so that $\sin(2\tilde{\phi}_{p,\tau})=\cos(2\tilde{\phi}_{\tau})$, and we arrive at a self-dual sine-Gordon model
\begin{widetext}
 \begin{align}
\label{H_eff_self-dual}
 &H^\sigma_{eff}=\int dx \left\{\frac{u_{p}}{2\pi}\left[\frac{1}{\tilde{K}_{p}}(\partial_x\tilde{\phi}_{\tau})^2+
 \tilde{K}_{p}(\partial_x\tilde{\theta}_{\tau})^2\right]+g\cos(\tilde{\theta}_{\tau})+\tilde{g}_P(\sigma,\tau)\cos(2\tilde{\phi}_{\tau})\right\}.
 \end{align}
 \end{widetext}
This is a special case of a series of models reviewed in Ref. [\onlinecite{Lech}]. In our case the choice $K\sim1/\sqrt{2}$ dictates $K_{p}\sim1$ and hence $\tilde{K}_{p}\sim1/2$, i.e. the quadratic part of the model is at the Heisenberg point, which is invariant under spin rotations. Then, we use the following relations
\begin{align}
\label{eq:fictspin}
\cos(\tilde{\theta}_{\tau})\sim(-)^x\sigma_x,\nonumber\\
\cos(2\tilde{\phi}_{\tau})\sim(-)^x\sigma_z,
\end{align}
where the $\sigma_a$ operators are pauli matrices representing fictitious local spins.
%The double sine-Gordon model Eq (\ref{eq:DSG}) can be mapped onto a spin-chain model in a staggered magnetic field in the $z-x$ plane, an angle $\alpha$ from the $\hat{z}$ direction.
The resulting (fictitious) spin model is
\begin{align}
\label{eq:stag-sc}
&H^\sigma_{eff}=\sum_i[J\sigma_i\cdot\sigma_{i+1}+(-)^x\vec{B}\cdot\sigma_i],\nonumber\\
&\vec{B}=\tilde{g}_P(\sigma,\tau)\hat{z}+g\hat{x}.
\end{align}
Eq. (\ref{eq:stag-sc}) describes a spin-chain model in a staggered magnetic field in the $z-x$ plane, at an angle
\begin{align}
\label{eq:alpha}&\alpha_{\tau}=\arctan(g/\tilde{g}_P(\sigma,\tau))
 \end{align}
from the $\hat{z}$ direction. Now we rotate the coordinate system so that the field will be in the $\hat{z}$ direction:
\begin{align}
\hat{x}'=\cos\alpha_{\tau}\hat{x}-\sin\alpha_{\tau}\hat{z},\nonumber\\
\hat{z}'=\sin\alpha_{\tau}\hat{x}+\cos\alpha_{\tau}\hat{z}\; .
\label{eq:rot}
\end{align}
The $\sigma_a$ spins are then related to the rotated spins $\sigma'_a$ by
\begin{align}
\label{eq:rotspin}
\sigma_x=\sigma_x'\cos\alpha_{\tau}+\sigma_z'\sin\alpha_{\tau},\nonumber\\
\sigma_z=\sigma_z'\cos\alpha_{\tau}-\sigma_x'\sin\alpha_{\tau}.
\end{align}
After this rotation the model is mapped onto a spin-chain in a staggered magnetic field along the $\hat{z}$ direction, which in Bosonization gives a regular sine-Gordon model with rotated fields $\phi'_{\tau},\;\theta'_{\tau}$:
\begin{widetext}
\begin{align}
\label{eq:HSG}
&H^\sigma_{eff}=\int dx \left\{\frac{u'}{2\pi}\left[\frac{1}{K'}(\partial_x\f'_{\tau})^2+K'(\partial_x\theta'_{\tau})^2\right]+g'_{\tau}\cos(2\phi'_{\tau})\right\},\quad
g'_{\tau}\equiv\sqrt{\tilde{g}_P^2(\sigma,\tau)+g^2},\;u'=u_{p},\;K'=\tilde{K}_{p}=\frac{K_{p}}{2}.
\end{align}
\end{widetext}
The $\cos(2\f'_{\tau})$ term opens a gap $\Delta'$ and obtains a finite expectation value. This term is also the order parameter of this model. Recalling that $K'\sim 1/2$, the system is deep in the gapped phase where a semi-classical approximation is justified to evaluate $\Delta'$. A variational calculation (see, e.g., Ref. [\onlinecite{gia}] for details) yields
\begin{equation}
\label{eq:delta'} \Delta'\sim
u'\Lambda\left(\frac{K'g'_{\tau}}{u'\Lambda^2}\right)^{1/(2-K')}
\end{equation}
with $\Lambda\sim 1/a$. The expectation value of the order parameter is subsequently given by
\begin{align}
\label{eq:order} &\langle\cos2\f'_{\tau}\rangle\sim
-\left(\frac{g'_{\tau}}{u'\Lambda^2}\right)^{K'}\sim
-\left(\frac{\Delta'}{u'\Lambda}\right)^{(2-K')K'}\; .
\end{align}
Substituting $K'=1/2$, this yields
\begin{align}
\label{eq:order_final} &\Delta'\sim
u'\Lambda\left(\frac{g'_{\tau}}{u'\Lambda^2}\right)^{2/3},\quad\langle\cos2\f'_{\tau}\rangle\sim
-\left(\frac{g'_{\tau}}{u'\Lambda^2}\right)^{1/2}\; .
\end{align}
Note that generically $g'_{\tau}>g,\tilde{g}_P(\sigma,\tau)$ [Eq. (\ref{eq:HSG})]; rather than competing with each other, the two self-dual interaction terms in Eq. (\ref{H_eff_self-dual}) cooperate to form an ordered ground state which smoothly evolves upon tuning of the parameters, and there is no phase transition.

\begin{figure}[h]
\begin{center}
\includegraphics[height=175pt,width=250pt]{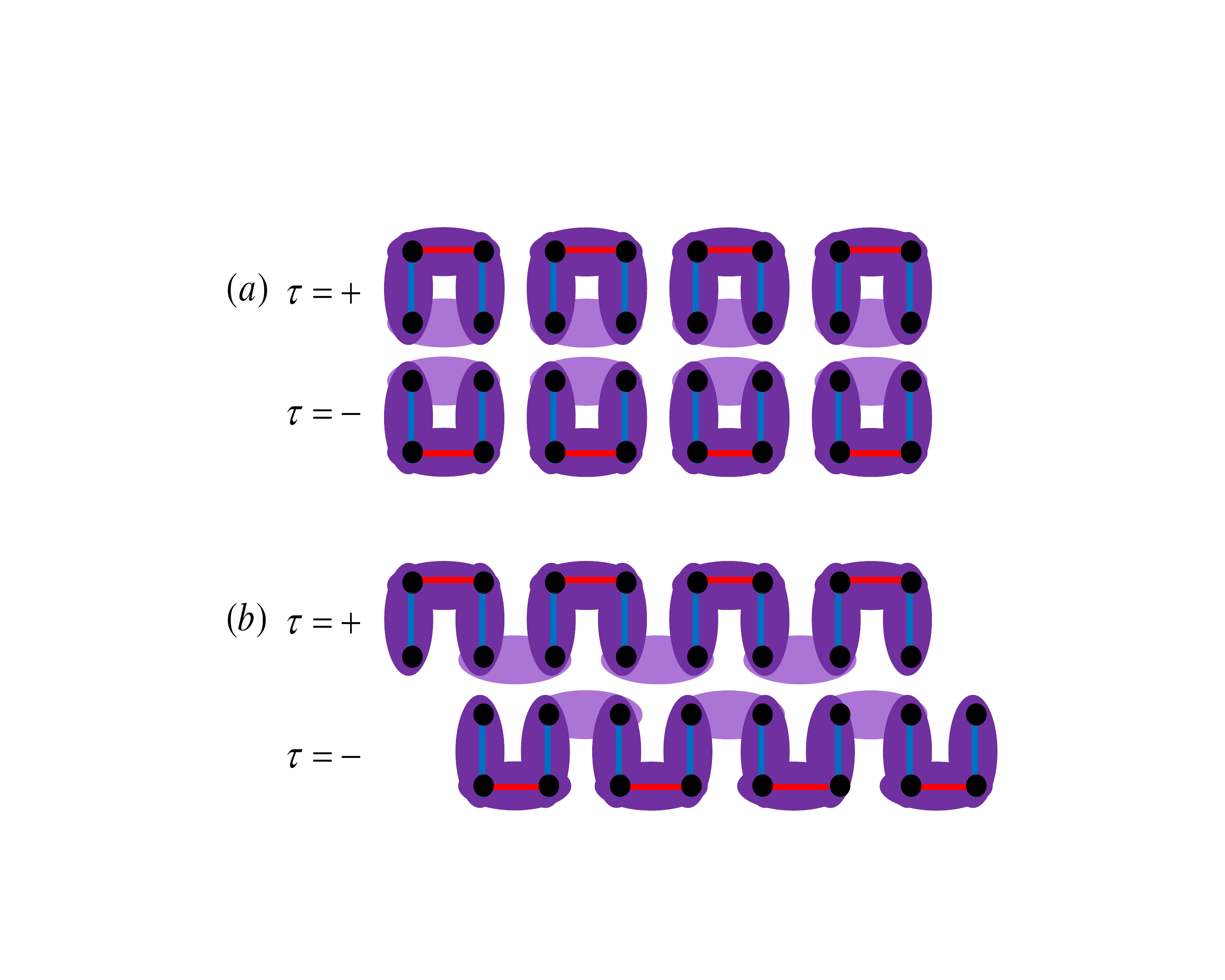}
\caption{(color online) Two possible types of plaquette order on the frustrated dimerized ladder: (a) closed  plaquettes corresponding to $\sigma=+$, (b) open plaquettes corresponding to $\sigma=-$. Dark purple ellipses represent strong dimers, and light purple ellipses weakened dimers. In each case, two distinct plaquette-ordered ground-state configurations emerge (corresponding to $\tau=\pm$), with spontaneously broken reflection symmetry.}
\label{fig:plaquettes}
\end{center}
\end{figure}

We next discuss the interpretation of the ordered state in terms of the physical spin system. Using Eqs. (\ref{eq:fictspin}), (\ref{eq:alpha}) and (\ref{eq:rotspin}) we express the order parameter field in terms of the fields $\tilde{\f}_{\tau}$ and $\tilde{\theta}_{\tau}$:
\begin{align}
\label{eq:cos2f'}
&\mathcal{P}_{\tau}\equiv\cos(2\f'_{\tau})=\cos\alpha_{\tau}\cos(2\tilde{\f}_{\tau})+\sin\alpha_{\tau}\cos(\tilde{\theta}_{\tau})\; ,\\
&\cos\alpha_{\tau}=\frac{\tilde{g}_P(\sigma,\tau)}{\sqrt{\tilde{g}_P^2(\sigma,\tau)+g^2}}\,,\quad \sin\alpha_{\tau}=\frac{g}{\sqrt{\tilde{g}_P^2(\sigma,\tau)+g^2}}.\nonumber
\end{align}
%Note that the sign  of $\cos\alpha$, and therefore the nature of the order parameter, depends on $\tau$ via $\tilde{g}_P(\sigma,\tau)$ [Eqs. (\ref{eq:stag-sc}), %(\ref{eq:alpha})] .
In both the CD ($\sigma=+$) and SD ($\sigma=-$) configurations, the ground state spontaneously breaks reflection symmetry across the ladder, with two distinct ground states (corresponding to $\tau=\pm$) of identical energies. To understand their physical significance, recall that $\cos(2\tilde{\f}_{\tau})$  and $\cos(\tilde{\theta}_{\tau})$ create longitudinal (on legs 1 and 2) and transverse dimers, respectively.
The corresponding local dimer operators are (see Appendix B)
\begin{align}
\label{eq:dimer_def}
&\epsilon_{l_{1}}\equiv S_{j,1}^+S_{j+1,1}^--S_{j,1}^-S_{j+1,1}^+\sim O_f^{\frac{1-\tau}{2}}\cos(2\tilde{\f}_{\tau}),\nonumber\\
&\epsilon_{l_{2}}\equiv S_{j,2}^+S_{j+1,2}^--S_{j,2}^-S_{j+1,2}^+\sim O_f^{\frac{1+\tau}{2}}\cos(2\tilde{\f}_{\tau}),\nonumber\\
&\epsilon_t\equiv S_{j,1}^+S_{j,2}^-+S_{j,1}^-S_{j,2}^+\sim \cos(\tilde{\theta}_{\tau})\nonumber\\
\end{align}
where $O_f$ is given by Eq. (\ref{eq:Of}).
Hence Eq. (\ref{eq:cos2f'}) implies that the order parameter is an entangled superposition of longitudinal and transverse dimers on plaquettes of four spins, i.e. a resonating valence bond within the plaquette; since $O_f<1$, for $\tau=+1$ ($\tau=-1$) the dimer operator on leg 1 (2) has a larger overlap with $\mathcal{P}_{\tau}$. The ground state is a crystal of such plaquettes, as illustrated in Fig. \ref{fig:plaquettes}. Since the dimers on chains 1,2 have two possible configurations, CD and SD, the plaquettes are also of two distinct types:
closed and open rectangular plaquettes, corresponding to the CD and SD states respectively.
%(see Fig. \ref{fig:plaquettes}).
The open rectangular plaquette order is relatively fragile, and under extreme conditions where $g'_{\tau}=0$, a gapless liquid state can be recovered. In comparison, the closed rectangle order is more robust and is lower in energy for given strength of the exchange interactions.

The long range order of dimers on distant plaquettes is reflected by the behavior of dimer-dimer correlation functions, which do not decay with increasing distance. We define
\begin{align}
\label{chimunu}
&\chi_{ab}(x,t)\equiv\langle\epsilon_a(x,t)\epsilon_b(0,0)\rangle,\;(a,b=l_1,l_2,t)
\end{align}
where $\epsilon_a$ are given by Eq. (\ref{eq:dimer_def}).
For $T\ll \Delta'$ and $x\gg\xi$, where $\xi=u'/\Delta'$ is the correlation length, these are approximated by constant asymptotic values:
 \begin{align}
 &\chi_{l_{\mu}l_{\nu}}(x\gg \xi)\cong O_f^{N_{\mu\nu}^{\tau}}\cos^2(\alpha_{\tau})(\Lambda\xi)^{-2K'},\nonumber\\
 &\chi_{tt}(x\gg \xi)\cong\sin^2(\alpha_{\tau})(\Lambda\xi)^{-2K'},\\
 &\chi_{l_{\nu}t}(x\gg \xi)\cong O_f^{N_{\nu\nu}^{\tau}/2}\frac{1}{2}\sin(2\alpha_{\tau})(\Lambda\xi)^{-2K'}\nonumber
 \end{align}
where $N_{\mu\nu}^{\tau}$ are given by
\begin{align}
&N_{11}^{\tau}=1-\tau,\;N_{22}^{\tau}=1+\tau,\;N_{12}^{\tau}=N_{21}^{\tau}=1
\end{align}
(see details of the calculation in Appendix \ref{app:corr}).
In particular, the long-range nature of $\chi_{l_{\nu}t}$, describing the correlation between a longitudinal and a transverse dimer, indicates the entanglement between two types of dimers within a plaquette, which is a consequence of the order parameter ${\mathcal P_\tau}$ being a superposition of longitudinal and transverse dimers [Eq. (\ref{eq:cos2f'})]. In the limit cases where either $g_P$ or $g$ vanishes, $\chi_{l_{\nu}t}=0$ and one recovers the rung or leg dimer VBC states, respectively.

\subsection{Phase Transition From VBC to Plaquette Order}

The calculation presented in subsection A suggests that away from the NT quantum critical point, and particularly for sufficient XXZ anisotropy of the rung coupling, the competition between the transverse and longitudinal dimerization terms may give rise to an ordered state of Plaquette dimers. However, since the leg dimerization term is strongly relevant, as long as $g_P$ is still relatively large the ground state will be dominated by the leg dimerization term, and a VBC state (as depicted in Fig. \ref{fig:SDCD}) will likely be favorable. This is especially notable in the case of the SD configuration ($\sigma=-$), where the Plaquette order is partially frustrated. When $g$ and $g_P$ are comparable, a first order transition from the VBC order to the plaquette order may occur, tuned by the ratio of $g$ and $g_P$. The transition line in the phase diagram, given in terms of the parameters $g$ and $g_P$, can be derived from energy considerations. To this end, we calculate the gain in energy for each phase to form, and compare them. The energy gain of a massive phase due to the ordering of a relevant operator is given by
\begin{align}
\label{eq:deltaE_def}
&\delta E\sim -\frac{\Delta^2}{E_0}
\end{align}
where $\Delta$ is the gap, and $E_0=u\Lambda$ (with $u$ a typical velocity) the high energy cutoff.
Similarly to the derivation of Eq. (\ref{eq:delta'}) for the gap in the plaquette ordered state, we employ
a variational approach to evaluate the gap opened by all relevant operators in terms of the parameters $g_P$ and $g^z$. This gives
\begin{align}
\label{eq:Deltas}
\Delta^z\sim u_s\Lambda\left(\frac{K_sg^z}{u_s\Lambda^2}\right)^{\frac{1}{2-2K_s}},\nonumber\\
\Delta_P\sim
u\Lambda\left(\frac{Kg_P}{u\Lambda^2}\right)^{\frac{1}{2-K}},
\end{align}
where $\Delta^z$  is the gap opened by $g^z\cos(\sqrt{8}\f_s)$ of Eq. (\ref{eq:HNT}) and $\Delta_P$ the gap opened
by the original leg dimerization term $H_P^\sigma$ [Eq. (\ref{eq:HP})].
Using these expressions we can calculate the gain in energy for the competing phases due to these operators. Forming a plaquette order will benefit the energy due to the gap $\Delta'$ and the energy due to the gap $\Delta^z$. Forming a VBC state will benefit the energy due to the gap opened by the leg dimerization, that is, twice (counted once for each chain) the energy gain from $\Delta_P$. Therefore we obtain the overall gain in energy for the competing phases to form:
\begin{align}
\label{eq:deltaE}
&\delta E_{plaq}=-(\Delta')^2/\Lambda u'-(\Delta^z)^2/\Lambda u_s\; ,\nonumber\\
&\delta E_{VBC}=-2(\Delta_P)^2/\Lambda u\; .
\end{align}
A transition between VBC order and plaquette order occurs when $\delta E_{plaq}=\delta E_{VBC}$. Using Eqs. (\ref{eq:HSG}), (\ref{eq:delta'}), (\ref{eq:Deltas}) and (\ref{eq:deltaE}) we plot a phase diagram for the transition from VBC order to plaquette order as a function of the strength of the rung dimerization $g$ and the leg dimerization $g_P$ for constant $g^z$. The result is presented in Fig. \ref{fig:phase diagram}.

\begin{figure}[h]
\begin{center}
\includegraphics[height=175pt,width=250pt]{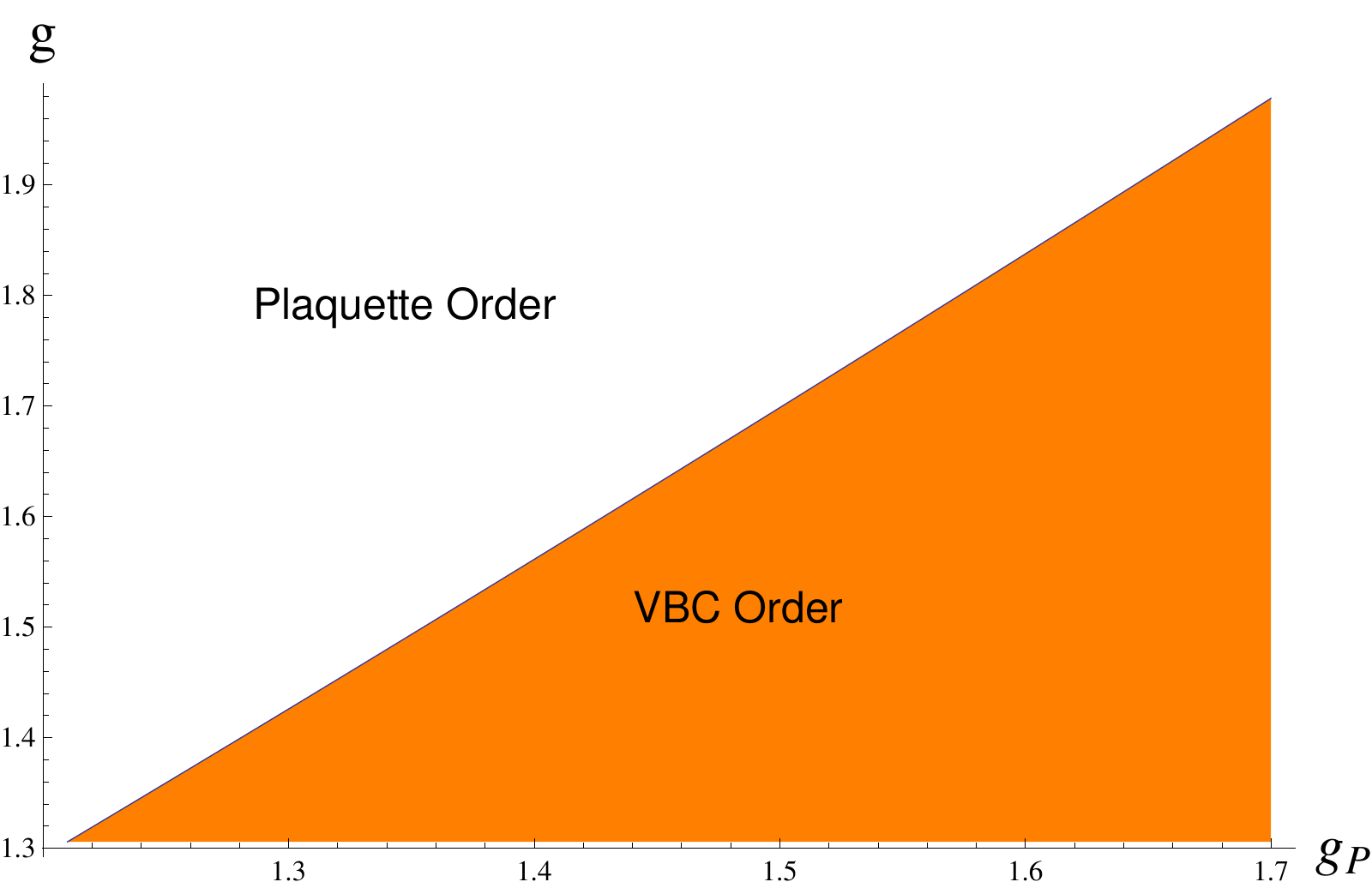}
\caption{(color online) $g-g_P$ phase diagram in arbitrary units
for fixed $g^z=1.5$, $\sigma=-$,
$K=1/\sqrt{2},\;\gamma=0.1,\;u=6.15$ and $\Lambda=\pi$. The phase
boundary line denotes a first order transition.} \label{fig:phase
diagram}
\end{center}
\end{figure}

\section{summary and discussion}
\label{sec:summary}
We studied a model for a dimerized frustrated ladder, namely a two-leg ladder version of the anisotropic NT model \cite{ners-2003} in the presence of dimerization on the legs.
Two types of dimerized patterns were considered: columnar dimers (CD) and staggered dimers (SD), which are respectively even and odd under reflection across the ladder (see Fig. \ref{fig:SDCD}). The effect of rung exchange interactions ($\Jper,\;J_d$) on the two configurations is distinct; for instance, effectively antiferromagnetic rung coupling ($\Jper-2J_d>0$) strengthens the ordering due to a CD instability while it frustrates the SD configuration (and the reverse for $\Jper-2J_d<0$). We particularly focus on the case of an intermediate anisotropy on the legs of the ladder where the Luttinger parameter $K\approx 1/\sqrt{2}$ [i.e. $\Jpar^z/\Jpar\approx 0.6$ - see Eq. (\ref{uK_def})], in which the leg dimerization terms and the rung interactions are equally relevant.
By mapping the resulting effective model to a spin-chain in a staggered magnetic field,
we found that the interplay between these interactions tends to form a ``Plaquette-ordered" phase: a crystal of resonating valence bonds plaquettes where reflection symmetry across the ladder is spontaneously broken (see Fig. \ref{fig:plaquettes}). The order parameter in this phase is a coherent superposition of longitudinal and transverse dimers, hence all types of dimer-dimer correlations are long-range.

The analysis leading to the above result relies on a mean-field approximation, justified when the rung exchange is tuned far enough from the NT quantum critical point $\Jper=2J_d$. The resulting gap to excitations is smaller in one of the dimerized configurations (e.g., for $(\Jper-2J_d)>0$ it is the SD configuration), and can even be tuned to zero for an extreme limit of the parameters. Under these extreme conditions, one apparently expects the formation of a gapless, Luttinger liquid mode (which can be interpreted as spin-$1/2$ chain meandering between the two legs of the ladder). Note that for more generic parameters, quantum fluctuations not accounted for in our low-energy approximations might also soften the gap: these introduce dynamics of the isospin auxiliary field $\hat{\tau}$, and consequently drive a transition of the Ising type to a liquid-like disordered phase with restored reflection symmetry. However, typically a gapless liquid state is unstable to other forms of order. In particular, for sufficiently strong dimerization (of either the CD or SD type), the Plaquette-ordered phase always gives way to the VBC state (i.e. a dimer-crystal of the corresponding structure) via a {\it first order} transition. A typical phase diagram is depicted in Fig. \ref{fig:phase diagram}.

It should be noted that while the analysis presented in the previous sections, which has focused on a special value of the leg-anisotropy ($K=1/\sqrt{2}$) allowing the exact mappings to free Fermions and the Heisenberg chain in a staggered field, the conclusions are more general. The formation of a Plaquette ordering essentially arises from the interplay of two highly relevant dimerization interactions, when their gap scales (as calculated for each interaction independently) are comparable. Deviations from $K=1/\sqrt{2}$ will thus lead to quantitative rather than qualitative corrections of our main results.

As a final remark, it is suggestive that our findings for the two-leg ladder version of the NT model are the key to understanding the behavior in physical realizations of the full-fledged 2D model such as the compound NOCuNO studied in Ref. [\onlinecite{vasiliev-2010}]. Similarly to the CD and SD instabilities introduced in this paper, in a multi-chain system a variety of lattice distortions generated by the softening of certain phonon modes may occur. This is especially expected in the presence of strong spin-phonon coupling. The resulting interplay between leg and rung dimerization interactions may give rise to various ordered states involving a superposition of transverse and longitudinal dimers, as generalizations of the Plaquette-ordered phase discussed in this paper. Moreover, several distinct broken symmetry states with identical or comparable energy may compete. As a consequence, one generically expects the formation of domains with different ordered spin-gapped configurations. The boundaries between domains can potentially support gapless liquid of spinons. This would be manifested as a partial contribution of gapless spinons to thermodynamic coefficients, as observed in the experiment \cite{vasiliev-2010}.

\acknowledgements

We thank L. Balents, R. Chandra, D. Podolsky, R. Santos and A. Tsvelik, and especially A. Vasiliev and O. Volkova for illuminating discussions. E. S. is grateful to the hospitality of the Aspen Center for Physics (NSF
grant 1066293) and to the Simons Foundation. This
work was supported by the Israel Science Foundation (ISF) grant 599/10.

\appendix
\section{Refermionization of $H_f$ and Mean-Field Approximation}
\label{app:referm}
Below we derive the mapping of $H_f$ [Eq. (\ref{eq:Hpf})] to gapped free Fermions, and calculate expectation values of operators in the $f$ sector using the refermionized version of these operators.
To this end, we employ Eq. (\ref{eq:psi}) (with $\f_{\nu},\theta_{\nu}$ replaced by $\f_f,\theta_f$). For $K_f=1$, the first term in Eq. (\ref{eq:Hpf}) reduces to a kinetic energy
\begin{align}
H_f^K=-iu_f\int\,dx\left\{\psi_R^\dagger\partial_x\psi_R-\psi_L^\dagger\partial_x\psi_L\right\}
\end{align}
and the second term is given by Eq. (\ref{eq:cosphi_ferm}).
Transforming to momentum space we thus obtain
\begin{widetext}
\begin{align}
&H_f=\sum_ku_fk(c_{R,k}^{\dagger}c_{R,k}-c_{L,k}^{\dagger}c_{L,k})+\Delta\sum_k(c_{R,k}^{\dagger}c_{L,k}+c_{L,k}^{\dagger}c_{R,k}),\nonumber\\
&\Delta\equiv\pi ag^z.
\end{align}
\end{widetext}
This Hamiltonian can be diagonalized by a Bogoliubov transformation\cite{gia}
\begin{align}
c_{+,k}^{\dagger}=\alpha_k c_{R,k}^{\dagger}+\beta_k c_{L,k}^{\dagger},\nonumber\\
c_{-,k}^{\dagger}=-\beta_k c_{R,k}^{\dagger}+\alpha_k c_{L,k}^{\dagger},
\label{eq:bog}
\end{align}
with
\begin{align}
\alpha_k=\frac{1}{\sqrt{2}}\left[1+\frac{u_fk}{\sqrt{(u_fk)^2+\Delta^2}}\right]^{1/2},\nonumber\\
\beta_k=\frac{1}{\sqrt{2}}\left[1-\frac{u_fk}{\sqrt{(u_fk)^2+\Delta^2}}\right]^{1/2},
\end{align}
after which $H_f$ becomes
\begin{align}
&H_f=\sum_k\sum_{\nu=\pm}E_{\nu,k}c_{\nu,k}^{\dagger}c_{\nu,k},\nonumber\\
&E_{\pm,k}=\pm\sqrt{(u_fk)^2+\Delta^2}.
\end{align}
For low $T\ll\Delta$, this justifies a mean-field approximation where we replace $\cos2\phi_f$ as well as $\partial_x\phi_f,\;\partial_x\theta_f$ and $\sin2\phi_f$ in $H_{pf}^\sigma$  by their expectation values. In terms of Fermionic fields these operators give by
\begin{align}
\sin2\phi_f=-i\pi a(\psi_R^{\dagger}\psi_L-H.c),\nonumber\\
\partial_x\phi_f=-\pi(\psi_R^{\dagger}\psi_R+\psi_L^{\dagger}\psi_L),\nonumber\\
\partial_x\theta_f=-\pi(\psi_R^{\dagger}\psi_R-\psi_L^{\dagger}\psi_L).
\end{align}
Fourier transforming and using Eq. (\ref{eq:bog}), we get (to leading order in $\frac{\Delta a}{u_f}$)
\begin{widetext}
\begin{align}
\label{eq:expect}
&\langle\sin2\phi_f\rangle=0,\;\langle\partial_x\phi_f\rangle=0,\langle\partial_x\theta_f\rangle=0,\nonumber\\
&O_f\equiv\langle\cos2\phi_f\rangle=\pi a\langle\psi_R^{\dagger}\psi_L+\psi_L^{\dagger}\psi_R\rangle=\sum_k\{\alpha_k\beta_k\langle c_{+,k}^{\dagger}c_{+,k}-c_{-,k}^{\dagger}c_{-,k}\rangle+
(\alpha_k^2-\beta_k^2)\langle c_{+,k}^{\dagger}c_{-,k}+c_{-,k}^{\dagger}c_{+,k}\rangle\},
\end{align}
\end{widetext}
which yields
\begin{align}
\label{eq:Ofapp}
&O_f\sim-\frac{\vert\Delta\vert}{u_f/a}\ln\left[\frac{u_f/a}{\vert\Delta\vert}\right].
\end{align}
Here we have used $\langle c_{\mu,k}^{\dagger}c_{\nu,k'}\rangle=\delta_{\mu\nu}\delta_{k,k'}f_{\mu,k}$ (for $\mu,\nu=\pm$), with $f_{\pm,k}=(1+e^{E_{\pm,k}/T})^{-1}$ the Fermi distribution function, approximated by $f_-\approx1,\;f_+\approx0$ for $T\ll\Delta$. Substituting back into Eq. (\ref{eq:Hpf}), this yields
the effective Hamiltonian Eq. (\ref{eq:Hp}).

\section{Dimer Correlation Functions}
\label{app:corr}
We are interested in the correlations between dimers of spins in the Plaquette ordered state. To this end, we define the local dimerization operators (at $x=ja$)
\begin{align}
\label{eq:dimer_def-app}
&\epsilon_{l_{\nu}}=S_{j,\nu}^+S_{j+1,\nu}^--S_{j,\nu}^-S_{j+1,\nu}^+=2i \sin(2\f_{\nu})\quad(\nu=1,2),\nonumber\\
&\epsilon_t=S_{j,1}^+S_{j,2}^-+S_{j,1}^-S_{j,2}^+=2\cos(\sqrt{2}\theta_a),
\end{align}
where the indices $l_{1/2},t$ stand for longitudinal (on chain 1 or 2) and transverse dimers respectively; note that in the expression for $\epsilon_{l_1}$, the site index $j$ is even while in $\epsilon_{l_2}$ it is even (odd) for $\sigma=+$ ($\sigma=-$). Recasting the Bosonic fields $\phi_\nu$ and $\theta_a$ in terms of the fields defined via the transformation Eq. (\ref{eq:nonunitary}) [and subsequently in terms of $\tilde{\phi}_{\tau}$, $\tilde{\theta}_{\tau}$ appearing in Eq. (\ref{H_eff_self-dual})],
after using the mean-field result Eq. (\ref{eq:expect}) we get
\begin{align}
\label{eq:nu2tau}
\sin(2\f_1)\sim O_f^{\frac{1-\tau}{2}}\cos(2\tilde{\f}_{\tau}),\nonumber\\
\sin(2\f_2)\sim O_f^{\frac{1+\tau}{2}}\cos(2\tilde{\f}_{\tau}),\nonumber\\
\cos(\sqrt{2}\theta_a)=\cos(\tilde{\theta}_{\tau}) .
\end{align}
Employing the mapping to fictitious spins [Eq. (\ref{eq:fictspin})], a similar mapping of $\cos(2\phi'_{\tau})$ and $\cos(\theta'_{\tau})$ to $\sigma'_x$, $\sigma'_z$ and the rotation Eq. (\ref{eq:rotspin}), one obtains
\begin{align}
\label{eq:epslt1}
\cos(2\tilde{\f}_{\tau})\sim\cos(2\phi'_{\tau})\cos\alpha_{\tau}-\cos(\theta'_{\tau})\sin\alpha_{\tau},\nonumber\\
\cos(\tilde{\theta}_{\tau})\sim\cos(\theta'_{\tau})\cos\alpha_{\tau}+\cos(2\phi'_{\tau})\sin\alpha_{\tau}.
\end{align}
The correlation functions between dimers are defined as
\begin{align}
\label{eq:chimunu}
&\chi_{\mu\nu}(x,t)=\langle\epsilon_{\mu}(x,t)\epsilon_{\nu}(0,0)\rangle,\;(\mu,\nu=l_1,l_2,t).
\end{align}
Using Eqs. (\ref{eq:dimer_def-app}), (\ref{eq:nu2tau}) and (\ref{eq:epslt1}) we thus obtain expressions for $\chi_{\mu\nu}(x,t)$ in terms of correlation functions of two types of operators: $\cos(2\phi'_{\tau})$ and $\cos(\theta'_{\tau})$.

In order to calculate correlation functions of the sine-Gordon model, we use the fact that in the gapped phase the cosine term can be expanded around the average value of $\phi'_{\tau}$, so that $\cos(2\phi'_{\tau})\cong 2(\phi'_{\tau}-\pi/2)^2-1$. Then, changing to a new field $\varphi_{\tau}=\phi'_{\tau}-\pi/2$, we arrive at a quadratic Hamiltonian for the massive field $\varphi_{\tau}$
\begin{align}
H=\frac{u'}{2\pi}\int
dx\left[\frac{1}{K'}(\partial_x\varphi_{\tau})^2+K'(\partial_x\theta'_{\tau})^2+\frac{(\Delta')^2}{K'(u')^2}\varphi_{\tau}^2\right],
\end{align}
with the gap $\Delta'$ given by Eq. (\ref{eq:delta'}).
The correlation functions of a Gaussian theory can be readily calculated using the methods shown in appendix C of Ref. [\onlinecite{gia}]. The correlations of the form $\langle\cos(\theta'_{\tau}(x,t))\cos(\theta'_{\tau}(0))\rangle$ decay exponentially  because  due to the uncertainty principle, when $\f'_{\tau}$ is ordered, $\theta'_{\tau}$ fluctuates. Therefore, at $T\ll\Delta'$ the only contributions to $\chi_{\mu\nu}(x,t)$ arise from the correlation function
\begin{equation}
\label{eq:C_varphi_def}
C_{\varphi}(\vec{r})\equiv\langle\cos(2\varphi_{\tau}(\vec{r})+\pi)\cos(2\varphi_{\tau}(0)+\pi)\rangle\sim e^{-2K'G_{\varphi}(\vec{r})}
\end{equation}
where the propagator $G_{\varphi}(\vec{r})\equiv\langle(\varphi_{\tau}(\vec{r})-\varphi_{\tau}(0))^2\rangle$ is given by
\begin{align}
G_{\varphi}(\vec{r})=\frac{1}{\beta\Omega}\sum_{\vec{q}}[1-\cos(\vec{q}\cdot\vec{r})]\frac{2\pi u'}{\omega_n^2+(u'k)^2+(\Delta')^2};
\end{align}
here $\vec{r}=(x,u'\tau)$ with $\tau$ the imaginary time, $\vec{q}=(k,\omega_n/u')$, $\beta=1/T$ and $\Omega$ the length. In the limit $\beta,\Omega\rightarrow\infty$, the sum can be transformed into an integral and one obtains
%\begin{widetext}
 \begin{align}
 G_{\varphi}(\vec{r})=\frac{ u'}{2\pi}\int_0^{\Lambda}qdq\int_0^{2\pi}d\theta_q\frac{1-\cos(qr\cos(\theta_q))}{(u'q)^2+(\Delta')^2},
 \end{align}
%\end{widetext}
 where $\theta_q$ is the angle between $\vec{q}$ and $\vec{r}$, $q=|\vec{q}|$ and $r=|\vec{r}|$. The result for $G_{\varphi}(\vec{r})$ is
 \begin{align}
 &G_{\varphi}(r)=\ln(\Lambda\xi)- K_0(r/\xi)  ,\nonumber\\
 &\xi\equiv u'/\Delta'
 \end{align}
 where $K_0(z)$ is the modified Bessel function. Using the asymptotic and series expansions of $K_0(z)$ for large and small arguments
 \begin{align}
 &K_0(z\gg 1)\approx\sqrt{\frac{\pi}{z}}e^{-z},\quad K_0(z\ll 1 )\approx-\ln(z),
 \end{align}
  we obtain a result for $C_{\varphi}$ [Eq. (\ref{eq:C_varphi_def})] in the two limits:
 \begin{align}
 &C_{\varphi}(r\ll\xi)\cong(r\Lambda)^{-2K'},\quad C_{\varphi}(r\gg\xi)\cong (\Lambda\xi)^{-2K'}\; .
 \end{align}
Finally, employing Eqs. (\ref{eq:dimer_def-app}), (\ref{eq:nu2tau}) and (\ref{eq:epslt1})
this yields the correlation functions [Eq. (\ref{eq:chimunu})] in the limit $r\gg\xi$:
% \begin{align}
% &\chi_{ll}\cong\cos^2(\alpha_{\tau})(\Lambda\xi)^{-2K'},\;\chi_{lt}\cong\frac{1}{2}\sin(2\alpha_{\tau})(\Lambda\xi)^{-2K'},\nonumber\\
% &\chi_{tt}\cong\sin^2(\alpha_{\tau})(\Lambda\xi)^{-2K'}.
% \end{align}
 \begin{align}
 &\chi_{l_{\mu}l_{\nu}}(x\gg \xi)\cong O_f^{N_{\mu\nu}^{\tau}}\cos^2(\alpha_{\tau})(\Lambda\xi)^{-2K'},\nonumber\\
 &\chi_{tt}(x\gg \xi)\cong\sin^2(\alpha_{\tau})(\Lambda\xi)^{-2K'},\\
 &\chi_{l_{\nu}t}(x\gg \xi)\cong O_f^{N_{\nu\nu}^{\tau}/2}\frac{1}{2}\sin(2\alpha_{\tau})(\Lambda\xi)^{-2K'}\nonumber
 \end{align}
where $N_{\mu\nu}^{\tau}$ are given by
\begin{align}
&N_{11}^{\tau}=1-\tau,\;N_{22}^{\tau}=1+\tau,\;N_{12}^{\tau}=N_{21}^{\tau}=1
\end{align}


\begin{thebibliography}{99}
\bibitem{foot1}
Note that in this limit, our derivation does not strictly apply since the underlying assumption of weak rung-coupling approximation breaks down.
\end{thebibliography}
\end{document}